\documentclass[aps,prl,reprint,groupedaddress]{revtex4-2}

\usepackage{amsmath,amssymb,amsthm,amsxtra,overpic,bbm,bm,epsfig,subfigure}
\usepackage{hyperref}
\usepackage{mathrsfs}
\usepackage{enumitem}
\usepackage{graphicx}
\usepackage{comment}
\usepackage{epstopdf}
\usepackage{float}
\usepackage{multirow}
\usepackage{verbatim} 
\usepackage{color}
\usepackage{slashed}
\usepackage{multirow}
\usepackage{pstricks}
\usepackage{booktabs}
\usepackage{epstopdf}
\usepackage{bbding}
\usepackage{braket}
\usepackage{blkarray}

\newcommand{\prlsection}[2]{{\it\textbf{#1}{#2}}---}
\makeatletter
\newcommand*{\balancecolsandclearpage}{%
	\close@column@grid
	\cleardoublepage
	\twocolumngrid
}
\makeatother

\def \bal#1\eal  {\begin{align} #1 \end{align}}
\newcommand{\be} {\begin{equation}}
	\newcommand{\ee} {\end{equation}}

\begin{document}

\title{Spelling Out Leptonic CP Violation in the Language of Invariant Theory}
\author{Bingrong Yu}
\email{yubr@ihep.ac.cn}
\affiliation{
	Institute for High Energy Physics, Chinese Academy of Sciences, Beijing 100049, China}

\affiliation{School of Physical Sciences, University
	of Chinese Academy of Sciences, Beijing 100049, China
}

\author{Shun Zhou}
\email{zhoush@ihep.ac.cn (corresponding author)}
\affiliation{
	Institute for High Energy Physics, Chinese Academy of Sciences, Beijing 100049, China}

\affiliation{School of Physical Sciences, University
	of Chinese Academy of Sciences, Beijing 100049, China
}

\date{\today}

\begin{abstract}
In terms of flavor invariants, we establish the intimate connection between leptonic CP violation in the canonical seesaw model for neutrino masses and that in the seesaw effective field theory (SEFT). For the first time, we calculate the Hilbert series and explicitly construct the primary flavor invariants in the SEFT by considering both the dimension-five Weinberg operator ${\cal O}^{\alpha \beta}_5 = \overline{\ell^{}_{\alpha \rm L}} \widetilde{H} \widetilde{H}^{\rm T} \ell^{\rm C}_{\beta \rm L}$ and the dimension-six operator ${\cal O}^{\alpha \beta}_6 = \left(\overline{\ell^{}_{\alpha \rm L}} \widetilde{H}\right) {\rm i}\slashed{\partial}\left( \widetilde{H}^\dagger \ell^{}_{\beta \rm L}\right)$ at the tree-level matching. The inclusion of only the Wilson coefficients $C^{\alpha \beta}_5$ and $C^{\alpha \beta}_6$ already enables the SEFT to incorporate all physical information about the full seesaw model. Moreover, the minimal sufficient and necessary conditions for CP conservation both in the SEFT and in the full theory are clarified, and the matching between the flavor invariants in both theories is accomplished. Through the matching of flavor invariants, the CP asymmetries necessary for successful leptogenesis are directly linked to those in neutrino-neutrino and neutrino-antineutrino oscillations at low energies. Surprisingly, it is revealed that the precise measurements of $C^{\alpha \beta}_5$ and $C^{\alpha \beta}_6$ in low-energy experiments are powerful enough to probe the full seesaw model, including CP violation for cosmological matter-antimatter asymmetry.
\end{abstract}

\maketitle

\prlsection{Introduction}{.} The violation of charge-parity (CP) symmetry should have played a crucially important role in the dynamical generation of matter-antimatter asymmetry in our Universe~\cite{Sakharov:1967dj, Bodeker:2020ghk}. While CP violation has been discovered in the quark sector~\cite{Christenson:1964fg, KTeV:1999kad, BaBar:2001pki}, a number of ongoing and forthcoming long-baseline accelerator neutrino oscillation experiments~\cite{T2K:2018rhz, DUNE:2015lol, Hyper-KamiokandeProto-:2015xww, Hyper-Kamiokande:2016srs} aim to probe CP violation in the leptonic sector~\cite{Branco:2011zb}.

In the standard model (SM), it is well known that the CP-violating phase in the Cabibbo-Kobayashi-Maskawa (CKM) matrix~\cite{Kobayashi:1973fv}, appearing in the charged-current interaction of quarks, accounts for the phenomena of CP violation observed in the meson systems. Though the standard parametrization of the CKM matrix~\cite{ParticleDataGroup:2020ssz} in terms of three flavor mixing angles and one Dirac-type CP-violating phase is given in the physical basis and thus widely adopted in flavor physics, the observables should be independent of both flavor bases and the specific parametrization of flavor mixing matrix. In a series of papers~\cite{Jarlskog:1985ht, Jarlskog:1985cw, Jarlskog:1986mm}, Jarlskog was the first to construct a basis- and parametrization-independent quantity to characterize CP violation~\cite{Jarlskog:1985cw}, namely,
\begin{eqnarray}
	{\rm Det} \left\{ [H^{}_{\rm u}, H^{}_{\rm d}] \right\} = 2{\rm i} \Delta^{}_{uc} \Delta^{}_{ct} \Delta^{}_{tu} \Delta^{}_{ds} \Delta^{}_{sb} \Delta^{}_{bd} {\cal J} \; ,
	\label{eq:Jarlskog}
\end{eqnarray}
where $H^{}_{\rm u} \equiv M^{}_{\rm u} M^\dagger_{\rm u}$ and $H^{}_{\rm d} \equiv M^{}_{\rm d} M^\dagger_{\rm d}$ with $M^{}_{\rm u}$ and $M^{}_{\rm d}$ being the up- and down-type quark mass matrices, respectively. In Eq.~(\ref{eq:Jarlskog}), $\Delta^{}_{qq^\prime} \equiv m^2_q - m^2_{q^\prime}$ denotes the quark mass-squared difference, and ${\cal J}$ is the Jarlskog rephasing invariant composed of the CKM matrix elements~\cite{Jarlskog:1985ht, Wu:1985ea}. Since $H^{}_{\rm u}$ and $H^{}_{\rm d}$ transform adjointly under the unitary transformations in the quark flavor basis, the determinant of their commutator is a flavor invariant. The vanishing of such a flavor invariant serves as the necessary and sufficient condition for CP conservation in the SM.

The construction of flavor invariants that are odd under the CP transformation has been generalized to an arbitrary number of generations of fermions in the SM in Ref.~\cite{Bernabeu:1986fc}, and to the leptonic sector with massive Majorana neutrinos~\cite{Branco:1986gr, Branco:2001pq, Branco:2006ce}. The minimal number of sufficient and necessary conditions for CP conservation in the presence of massive Majorana neutrinos and lepton mass degeneracy have been studied in Refs.~\cite{Yu:2019ihs, Yu:2020gre, Yu:2020xyy}. Only in Ref.~\cite{Jenkins:2009dy} was it first pointed out that the Hilbert series (HS) in the invariant theory is a powerful mathematical tool for a systematic study of flavor invariants and their relationships with physical parameters in flavor physics. Moreover, the plethystic program~\cite{Benvenuti:2006qr} has been implemented in Ref.~\cite{Hanany:2010vu} to calculate the HS for the ring of invariants through the Molien-Weyl (MW) formula~\cite{Molien1897,Weyl1926}. It has been clarified in Ref.~\cite{Jenkins:2009dy} that the number of primary invariants is equal to that of independent physical parameters in the theory, whereas all the flavor invariants can be expressed as the polynomials of the basic invariants in the generating set.

In the type-I seesaw model~\cite{Minkowski:1977sc, Yanagida:1979as, GellMann:1980vs, Glashow:1979nm, Mohapatra:1979ia} and its low-energy effective theory with only the dimension-five Weinberg operator~\cite{Weinberg:1979sa}, the basic flavor invariants have been partly investigated~\cite{Jenkins:2009dy} and their renormalization-group equations are calculated in Ref.~\cite{Wang:2021wdq}. In the minimal seesaw model with two right-handed (RH) neutrinos, all the basic flavor invariants have been explicitly constructed and connected to the flavor invariants in the effective theory by a proper matching procedure~\cite{Yu:2021cco}. Recently, the CP-odd flavor invariants have been examined in Ref.~\cite{Bonnefoy:2021tbt} in the Standard Model effective field theory (SMEFT) with non-renormalizable operators of mass dimension up to six~\cite{Buchmuller:1985jz, Grzadkowski:2010es, Brivio:2017vri}. However, the flavor mixing and CP violation in the leptonic sector have been switched off in Ref.~\cite{Bonnefoy:2021tbt}, as the Weinberg operator is ignored and thus no lepton flavor mixing occurs.

In this letter, we explore the flavor invariants in the type-I seesaw model and those in the seesaw effective field theory (SEFT) at the tree-level matching, where both the Weinberg operator ${\cal O}^{\alpha \beta}_5 = \overline{\ell^{}_{\alpha \rm L}} \widetilde{H} \widetilde{H}^{\rm T} \ell^{\rm C}_{\beta \rm L}$ and the dimension-six operator ${\cal O}^{\alpha \beta}_6 = \left(\overline{\ell^{}_{\alpha \rm L}} \widetilde{H}\right) {\rm i}\slashed{\partial}\left( \widetilde{H}^\dagger \ell^{}_{\beta \rm L}\right)$~\cite{Broncano:2002rw,Broncano:2003fq} are present (here $\tilde{H}\equiv {\rm i}\sigma_2^{}H_{}^{*}$ denotes the Higgs doublet). In the language of invariant theory, we are able to draw a number of interesting conclusions. First, the inclusion of only two Wilson coefficients $C^{\alpha \beta}_5$ and $C^{\alpha \beta}_6$ in the SEFT reproduces the same number of physical parameters as in the full seesaw model. Second, in connection to the previous observation, we demonstrate that the absence of CP violation in the SEFT guarantees CP conservation in the full theory, and vice versa. The minimal sufficient and necessary conditions for CP conservation are given. In addition, we show that all physical parameters in the SEFT can be extracted using primary flavor invariants, so any low-energy physical observables can be expressed as functions of flavor invariants. Finally, the matching between the flavor invariants in the effective and full theories is accomplished. As a consequence, CP asymmetries necessary for a successful leptogenesis for cosmological matter-antimatter asymmetry~\cite{Fukugita:1986hr} can be directly related to those in neutrino-neutrino and neutrino-antineutrino oscillations at low energies.

\vspace{0.3cm}

\prlsection{Framework}{.} To accommodate nonzero neutrino masses, we work in the type-I seesaw model with $n$ RH neutrinos $N^{}_{\rm R}$. Apart from the SM Lagrangian, the RH neutrino part of the full theory is given by
\begin{eqnarray}
	\label{eq:full lagrangian}
	{\cal L} = \overline{N^{}_{\rm R}}{\rm i}\slashed{\partial} N^{}_{\rm R} - \left[\overline{\ell_{\rm L}^{}}Y_\nu^{}\tilde{H}N_{\rm R}^{} + \frac{1}{2}\overline{N_{\rm R}^{\rm C}}M_{\rm R}^{}N_{\rm R}^{}+{\rm h.c.} \right] \;, \quad
\end{eqnarray}
where ${\ell }_{\rm L}^{}$ stands for the left-handed lepton doublet. In Eq.~(\ref{eq:full lagrangian}), $Y_\nu^{}$ denotes the Dirac neutrino Yukawa coupling matrix and $M_{\rm R}^{}$ is the Majorana mass matrix of RH neutrinos. 

For the mass scale $\Lambda = {\cal O}(M^{}_{\rm R})$ of RH neutrinos much higher than the electroweak scale $v\approx 246\,{\rm GeV}$, the low-energy phenomena are described by the SEFT with
\begin{eqnarray}\label{eq:Left}
	\mathcal{L}^{}_{\rm SEFT} = \mathcal{L}^{}_{\rm SM} - \left[ \frac{C^{}_5}{2\Lambda} {\cal O}^{}_5 + {\rm h.c.} \right] + \frac{C^{}_6}{\Lambda^2} {\cal O}^{}_6 \; ,
\end{eqnarray}
where ${\cal L}^{}_{\rm SM}$ stands for the SM Lagrangian, and $\Lambda$ is the cutoff scale. At the tree-level matching, the relevant Wilson coefficients can be identified as
\begin{eqnarray}
	\label{eq:wilson coe}
	C_5^{}=-Y_\nu^{}Y_{\rm R}^{-1}Y_\nu^{\rm T}\;, \quad
	C_6^{}=Y_\nu^{} \left(Y_{\rm R}^{\dagger}Y_{\rm R}^{}\right)_{}^{-1}Y_\nu^\dagger\;,
\end{eqnarray}
with $Y_{\rm R}^{}\equiv M_{\rm R}^{}/\Lambda$. 
Taking account of the charged-lepton part from the SM, we consider the most general flavor-basis transformations in the leptonic sector
\begin{eqnarray}
	\label{eq:field trans}
	\ell_{\rm L}^{}\to U_{\rm L}^{}\ell_L^{}\;,\quad
	l_{\rm R}^{}\to V_{\rm R}l_{\rm R}^{}\;,\quad
	N_{\rm R}^{}\to U_{\rm R}^{}N_{\rm R}^{}\;,
\end{eqnarray}
where $l^{}_{\rm R}$ represents the RH charged-lepton fields, and $U_{\rm L}^{}, V_{\rm R} \in {\rm U}(m)$ and $U_{\rm R}^{} \in {\rm U}(n)$ are three arbitrary unitary matrices (for $m$ generations of lepton doublets and $n$ generations of RH neutrinos). Then Eq.~(\ref{eq:full lagrangian}) is unchanged if we treat the Yukawa coupling matrices as spurions, namely, taking them as spurious fields that transform as
\begin{eqnarray}
	\label{eq:Yukawa trans}
	Y_l^{} \to U_{\rm L}^{}Y_l^{}V_{\rm R}^\dagger\;,~~
	Y_\nu^{} \to  U_{\rm L}^{}Y_\nu^{}U_{\rm R}^\dagger\;, ~~
	Y_{\rm R}^{} \to  U_{\rm R}^* Y_{\rm R}^{}U_{\rm R}^\dagger\;, ~~
\end{eqnarray}
where $Y^{}_l$ is the charged-lepton Yukawa coupling matrix. At the matching scale, such transformations in the lepton flavor space in the full theory induce those of the Wilson coefficients in the SEFT, i.e.,
\begin{eqnarray}
	\label{eq:wilson coe trans}
	C_5^{}\to U_{\rm L}^{}C_5^{}U_{\rm L}^{\rm T}\;,\quad
	C_6^{}\to U_{\rm L}^{}C_6^{}U_{\rm L}^\dagger\;.
\end{eqnarray}
From Eq.~(\ref{eq:Yukawa trans}) and Eq.~(\ref{eq:wilson coe trans}) we can take the matrices $\left(X_l^{} \equiv Y^{}_l Y^\dagger_l, C_5^{},C_6^{}\right)$ in the flavor space as the building blocks for the flavor invariants in the SEFT with the symmetry group ${\rm U}(m)$, whereas $\left(Y_l^{},Y_\nu^{},Y_{\rm R}^{}\right)$ as the building blocks in the full seesaw model with the symmetry group ${\rm U}(m)\otimes {\rm U}(n)$. 

Throughout this letter, we use ${\cal I}_{abc}^{}$ to label the flavor invariant with the degrees $\left(a,b,c\right)$ of the building blocks $\left(X_l^{},C_5^{},C_6^{}\right)$ in the SEFT. Similarly, $I_{abc}^{}$ refers to the flavor invariant with the degrees $\left(a,b,c\right)$ of the building blocks $\left(Y_l^{},Y_\nu^{},Y_{\rm R}^{}\right)$ in the full seesaw model. Here $a,b,c$ are non-negative integers. By flavor invariants, we mean the \emph{polynomial} matrix invariants composed of building blocks that keep unchanged under flavor transformation.

\vspace{0.3cm}

\prlsection{Two-generation SEFT}{.} We begin with the case of only two generations of leptons. Although this is not realistic, it is very instructive for the study of the three-generation case. All the basic flavor invariants in both effective and full theories in the two-generation case can be explicitly constructed, and related to the physical observables in an apparent way.

As has been stated above, the HS is a powerful tool in studying the flavor invariants and the algebraic structure of the invariant ring. In the SEFT with two generations, using the MW formula, one can calculate the HS
\begin{eqnarray}
	\label{eq:HS eff 2g main}
	{\mathscr H}_{\rm SEFT}^{(2\rm g)}(q)=\frac{1+3q^4+2q^5+3q^6+q^{10}}{\left(1-q\right)^2\left(1-q^2\right)^4\left(1-q^3\right)^2\left(1-q^4\right)^2}\;,\quad
\end{eqnarray}
where $q$ is an arbitrary complex number that labels the degrees of the invariants. The denominator of the HS carries the information about the \emph{primary} invariants, i.e., those invariants that are algebraically independent. There are 10 factors in the denominator of the HS in Eq.~(\ref{eq:HS eff 2g main}), which means there are totally 10 primary flavor invariants in the invariant ring. The nontrivial point is that this number also equals the number of the independent physical parameters in the two-generation SEFT (i.e., 2 charged-lepton masses, 2 neutrino masses, 1 mixing angle and 1 phase in the leptonic flavor mixing matrix, 3 moduli and 1 phase in $C_6^{}$). As we will show below, all the 10 physical parameters can be extracted as the functions of 10 primary invariants.
\renewcommand\arraystretch{1.2}
\begin{table}[t!]
	\centering
	\begin{tabular}{l|c|c}
		\hline \hline
		flavor invariants &  degree & CP parity \\
		\hline \hline
		${\cal I}_{100}^{}\equiv {\rm Tr}\left(X_l^{}\right)\quad (*)$ &  1 & $+$ \\
		\hline
		${\cal I}_{001}^{}\equiv {\rm Tr}\left(C_6^{}\right)\quad (*)$ &  1 & $+$\\
		\hline
		${\cal I}_{200}^{}\equiv {\rm Tr}\left(X_l^2\right)\quad (*)$ &  2 & $+$\\
		\hline
		${\cal I}_{101}^{}\equiv {\rm Tr}\left(X_l^{}C_6^{}\right)$ & 2 & $+$\\
		\hline
		${\cal I}_{020}^{}\equiv {\rm Tr}\left(X_5^{}\right)\quad (*)$ & 2 & $+$\\
		\hline
		${\cal I}_{002}^{}\equiv {\rm Tr}\left(C_6^2\right)\quad (*)$ &  2 & $+$\\
		\hline
		${\cal I}_{120}^{}\equiv {\rm Tr}\left(X_l^{}X_5^{}\right)\quad (*)$ &  3 & $+$\\
		\hline
		${\cal I}_{021}^{}\equiv {\rm Tr}\left(C_6^{}X_5{}\right)\quad (*)$ & 3 & $+$\\
		\hline
		${\cal I}_{220}^{}\equiv {\rm Tr}\left(X_l{}G_{l5}^{}\right)\quad (*)$ & 4 & $+$\\
		\hline
		${\cal I}_{121}^{(1)}\equiv {\rm Tr}\left(G_{l5}^{}C_6^{}\right)$ & 4 & $+$\\
		\hline
		${\cal I}_{121}^{(2)}\equiv {\rm Im}\,{\rm Tr}\left(X_l^{}X_5{}C_6^{}\right)$ & 4 & $-$\\
		\hline
		${\cal I}_{040}^{}\equiv {\rm Tr}\left(X_5^{2}\right)\quad (*)$ & 4 & $+$\\
		\hline
		${\cal I}_{022}^{}\equiv {\rm Tr}\left(C_6^{}G_{56}^{}\right)\quad (*)$ & 4 & $+$\\
		\hline
		${\cal I}_{221}^{}\equiv {\rm Im}\,{\rm Tr}\left(X_l^{}G_{l5}^{}C_6^{}\right)$ & 5 & $-$\\
		\hline
		${\cal I}_{122}^{}\equiv {\rm Im}\,{\rm Tr}\left(C_6^{}G_{56}^{}X_l^{}\right)$ & 5 & $-$\\
		\hline
		${\cal I}_{240}^{}\equiv {\rm Im}\,{\rm Tr}\left(X_l^{}X_5^{}G_{l5}^{}\right)$ & 6 & $-$\\
		\hline
		${\cal I}_{141}^{}\equiv {\rm Im}\,{\rm Tr}\left(X_5^{}C_6^{}G_{l5}^{}\right)$ & 6 & $-$\\
		\hline
		${\cal I}_{042}^{}\equiv {\rm Im}\,{\rm Tr}\left(C_6^{}X_5^{}G_{56}^{}\right)$ & 6 & $-$\\
		\hline
		\hline
	\end{tabular}
	\vspace{0.5cm}
	\caption{Summary of the basic flavor invariants along with their degrees and CP parities in the case of two-generation leptons in the SEFT, where the subscripts of the invariants denote the degrees of $X_l^{}\equiv Y_l^{}Y_l^\dagger$, $C_5^{}$ and $C_6^{}$, respectively. We have also defined $X_5^{}\equiv C_5^{}C_5^\dagger$, $G_{l5}^{}\equiv C_5^{}X_l^*C_5^\dagger$ and $G_{56}^{}\equiv C_5^{}C_6^*C_5^\dagger$ that transform adjointly under the flavor transformation. There are in total 12 CP-even basic invariants and 6 CP-odd basic invariants. Note that the 10 primary invariants are labeled with ``$(*)$" in the first column.}
	\label{table:2g eff}
\end{table}
\renewcommand\arraystretch{1}

Although other invariants in the ring are not algebraically independent of the primary ones, not all of them can be written as the polynomials of the primary invariants. However, 
for the unitary groups under consideration,
one can always find a finite number of invariants, known as \emph{basic} invariants, such that any invariant in the ring can be decomposed as the polynomial of the basic invariants~\cite{Wang:2021wdq, Yu:2021cco, Sturmfels2008,DK2015}. In general the number of basic invariants is no smaller than that of primary invariants. This is because there may exist nontrivial polynomial identities among the basic invariants (i.e., the syzygies).

The construction of all the basic invariants can be accomplished by calculating the plethystic logarithm (PL) function of the HS
\begin{eqnarray}
	\label{eq:PL eff 2g main}
	{\rm PL}\left[{\mathscr H}_{\rm SEFT}^{(2\rm g)}(q)\right] &=& 2q+4q^2+2q^3+5q^4 \nonumber \\ &~& +2q^5+3q^6-6q^8-{\cal O}\left(q^9\right)\;,
\end{eqnarray}
whose leading positive terms encode the information about the numbers and degrees of the basic invariants~\cite{Benvenuti:2006qr}. As indicated by Eq.~(\ref{eq:PL eff 2g main}), there are totally 18 [obtained from the sum of all the coefficients in Eq.~(\ref{eq:PL eff 2g main}) until the first negative term] basic invariants in the ring: two of degree 1, four of degree 2, two of degree 3, five of degree 4, two of degree 5 and three of degree 6. Furthermore, we can explicitly construct all the basic flavor invariants in the two-generation SEFT, and the results are summarized in Table~\ref{table:2g eff}. The parities of basic flavor invariants under the CP transformation have been listed in the last column. The 18 basic invariants (12 CP-even and 6 CP-odd) in Table~\ref{table:2g eff} serve as the generators of the invariant ring in the sense that any flavor invariant can be written as the polynomial of them. For a systematic algorithm of decomposing an arbitrary invariant into the polynomial function of the basic invariants and finding out all the syzygies at a certain degree, see Appendix C of Ref.~\cite{Wang:2021wdq}.

However, those 18 basic flavor invariants in Table~\ref{table:2g eff} are not algebraically independent. As one can verify, there are 6 syzygies first appearing at degree 8, corresponding to the first negative term $-6q_{}^8$ in Eq.~(\ref{eq:PL eff 2g main}). Among them, four syzygies imply 4 linear relations among 6 CP-odd basic invariants and another two involve only CP-even invariants. This is in accordance with the fact that there are only $6-4=2$ independent phases in the two-generation case of the SEFT. 

In Table~\ref{table:2g eff}, ten primary flavor invariants are labeled by ``$(*)$". It can be shown that from them one can extract all the physical parameters in the two-generation SEFT (cf. Supplemental Materials). In this sense, the set of primary invariants is actually equivalent to that of independent physical parameters in the theory. Therefore, one can express any low-energy physical observables in an explicit and basis-independent form with only flavor invariants. In particular, any CP-violating observable ${\cal A}_{\rm CP}^{}$ can be written as~\cite{Yu:2022}
\begin{eqnarray}
	\label{eq:general CP asy}
{\cal A}_{\rm CP}^{}=\sum_j^{} {\cal F}_j^{}\left[{\cal I}_k^{\rm even}\right] {\cal I}_j^{\rm odd}\;,
\end{eqnarray}
where ${\cal I}_j^{\rm odd}$ refer to CP-odd basic flavor invariants, and ${\cal F}_j^{}\left[{\cal I}_k^{\rm even}\right]$ are some functions of only CP-even basic flavor invariants. Thus the vanishing of all CP-odd basic invariants in the ring ensures the absence of CP violation in the theory. We shall leave the proof of this general formula for Ref.~\cite{Yu:2022}. Instead we mention that CP asymmetries ${\cal A}_{\nu\nu}^{}$ in neutrino oscillations and those ${\cal A}_{\nu\bar{\nu}}^{}$ in neutrino-antineutrino oscillations~\cite{Xing:2013ty, Xing:2013woa, Wang:2021rsi} can indeed be cast in the form of Eq.~(\ref{eq:general CP asy}). After some lengthy calculations, we obtain ${\cal A}_{\nu\nu}^{}={\cal F}_{\nu\nu}^{}{\cal I}_{121}^{(2)}$ and ${\cal A}_{\nu\bar{\nu}}^{}={\cal F}_{\nu\bar{\nu}}^{}{\cal I}_{240}^{}$, where ${\cal F}_{\nu\nu}^{}$ and ${\cal F}_{\nu\bar{\nu}}^{}$ are functions of CP-even primary invariants while ${\cal I}_{121}^{(2)}$ and ${\cal I}_{240}^{}$ are two CP-odd basic invariants in Table~\ref{table:2g eff}.

Finally, we discuss the conditions for CP conservation. Though there are six CP-odd basic invariants in the ring, only two of them are algebraically independent due to the syzygies. On the other hand, there are two independent phases in the leptonic sector. Hence the \emph{minimal} conditions to guarantee CP conservation is the vanishing of only two CP-odd invariants. We find that the vanishing of ${\cal I}_{121}^{(2)}$ and ${\cal I}_{240}^{}$ is sufficient to this end. Therefore, CP asymmetries in neutrino oscillations and neutrino-antineutrino oscillations already contain all the information about CP violation at low energies.

\vspace{0.3cm}

\prlsection{Two-generation Seesaw}{.} In the full seesaw model, the building blocks transform in the flavor space as in Eq.~(\ref{eq:Yukawa trans}). Then the HS can be computed as~\cite{Jenkins:2009dy}
\begin{eqnarray*}
\label{eq:HS seesaw 2g}
{\mathscr H}_{\rm SS}^{(2\rm g)}(q)=\frac{1+q^6+3q^8+2q^{10}+3q^{12}+q^{14}+q^{20}}{\left(1-q^2\right)^3\left(1-q^4\right)^5\left(1-q^6\right)\left(1-q^{10}\right)}\;,
\end{eqnarray*}
which exhibits the algebraic structure of the flavor space in the full theory. We observe that the denominator of the HS in the full theory and that of Eq~(\ref{eq:HS eff 2g main}) have the same number of factors, implying that there are equal number of algebraically-independent invariants (i.e., primary invariants) in the flavor space of full theory and that of the SEFT. Given the fact that the number of primary invariants is equal to that of independent physical parameters, we reach the conclusion that inclusion of just one dimension-five and one dimension-six operator in the effective theory is already \emph{adequate} to incorporate all physical information about the full theory, including the source of CP violation~\cite{Broncano:2002rw, Broncano:2003fq,Antusch:2009gn}.

This point can be seen more clearly from the basic invariants. In the two-generation case, one can explicitly construct all the basic flavor invariants in the full theory, as listed in Table~\ref{table:2g seesaw}. To one's surprise, there are exactly equal number of CP-odd and CP-even basic invariants in Table~\ref{table:2g eff} and Table~\ref{table:2g seesaw}, namely, both are 6 and 12, respectively. Recalling that the basic invariants serve as the generators of the invariant ring, we conclude that the invariant ring in the SEFT and that in the full theory share the equal number of generators. One can establish a direct link between these two sets of generators by noticing that the building blocks $C_5^{}$ and $C_6^{}$ in the SEFT are related to the building blocks $Y_\nu^{}$ and $Y_{\rm R}^{}$ in full theory via Eq.~(\ref{eq:wilson coe}). Through a proper matching procedure~\cite{Yu:2021cco,Yu:2022}, we find \emph{all flavor invariants in the SEFT can be written as the rational functions of those in the full seesaw model.}

We have verified that all the 18 basic flavor invariants in the SEFT can be explicitly expressed as rational functions of the 18 basic flavor invariants in the full seesaw model. The complete set of matching conditions are given in Supplemental Materials. In particular, one can set up a \emph{one-to-one} correspondence between 6 CP-odd basic invariants in the SEFT and those in the full theory, namely,
\begin{widetext}
{\allowdisplaybreaks
	\begin{eqnarray}
		{\cal I}_{121}^{(2)}&=&\frac{2}{\left(I_{002}^2-I_{004}\right)^2}\left[I_{242}^{(2)}I_{022}^{}-I_{044}^{}I_{220}^{}+I_{262}^{}I_{002}^{}-I_{244}^{}I_{020}^{}\right]\;,\label{eq:odd1}\\
		{\cal I}_{221}^{}&=&\frac{2}{\left(I_{002}^2-I_{004}\right)^2}\left[I_{242}^{(2)}I_{222}^{}+I_{244}^{}I_{220}^{}+I_{462}^{}I_{002}^{}-I_{444}^{}I_{020}^{}\right]\;,\label{eq:odd2}\\
		{\cal I}_{122}^{}&=&\frac{2}{\left(I_{002}^2-I_{004}\right)^3}\left\{I_{242}^{(2)}\left[3I_{022}^2+2I_{040}^{}\left(I_{002}^2-I_{004}^{}\right)-4I_{020}^{}I_{002}^{}I_{022}^{}\right] +I_{244}^{}\left(3I_{020}^{}I_{022}^{}-2I_{042}^{}\right) \right.\nonumber\\
		&&\left.+I_{044}^{}\left(4I_{020}^{}I_{222}^{}-I_{220}^{}I_{022}^{}-2I_{242}^{(1)}\right)+I_{262}^{}\left[3I_{002}^{}I_{022}^{}-I_{020}^{}\left(I_{002}^2+3I_{004}^{}\right)\right]\right\} \; , \label{eq:odd3} \\
		{\cal I}_{240}^{}&=&\frac{1}{\left(I_{002}^2-I_{004}\right)^2}\left[3I_{242}^{(2)}\left(I_{022}^{}I_{220}^{}-I_{020}^{}I_{222}^{}\right)-I_{044}^{}I_{220}^2+I_{262}^{}\left(3I_{002}^{}I_{220}^{}-2I_{222}^{}\right) -2 I_{244}^{}I_{020}^{}I_{220}^{}\right.\nonumber\\
		&&\left.+I_{462}^{}\left(2I_{022}^{}-3I_{002}^{}I_{020}\right)+I_{444}^{}I_{020}^2\right]\;,\label{eq:odd4}\\
		{\cal I}_{141}^{}&=&\frac{2}{\left(I_{002}^2-I_{004}\right)^3}\left\{I_{242}^{(2)}I_{020}^{}I_{022}^2+I_{044}^{}I_{020}^{}\left(I_{022}^{}I_{220}^{}-2I_{242}^{(1)}\right) +I_{244}^{}I_{020}^{}\left(I_{020}^{}I_{022}-2I_{042}^{}\right) \right.\nonumber\\
		&&\left.+I_{262}^{}\left[I_{002}^{}I_{020}^{}I_{022}^{}+I_{040}^{}\left(I_{004}^{}-I_{002}^2\right)\right]\right\}\;,\label{eq:odd5}\\
		{\cal I}_{042}^{}&=&\frac{2}{\left(I_{002}^2-I_{004}\right)^3}\,I_{044}^{}\left(I_{020}^2-I_{040}^{}\right)_{}^2\;.\label{eq:odd6}
	\end{eqnarray}
}
\end{widetext}
Notice that Eqs.~(\ref{eq:odd1})-(\ref{eq:odd6}) form a system of \emph{linear} equations for the CP-odd invariants and the determinant of the coefficient matrix in Eqs.~(\ref{eq:odd1})-(\ref{eq:odd6}) turns out to be nonzero in general. This proves that the vanishing of all the CP-odd flavor invariants in the SEFT is equivalent to the vanishing of all CP-odd invariants in the full theory. Therefore, the absence of CP violation in the low-energy effective theory up to the order of ${\cal O}\left(1/\Lambda_{}^2\right)$ is equivalent to the CP conservation in the full seesaw model. Note that similar conclusion was also drawn in Ref.~\cite{Broncano:2003fq}, but without the language of invariant theory.

The matching conditions in Eqs.~(\ref{eq:odd1})-(\ref{eq:odd6}) are useful to build a bridge between the CP violation at low energies and
that at high energies. For example, if RH neutrino masses are strongly hierarchical, the (unflavored) CP asymmetry in the decay of the lightest RH neutrino can simply be written as ~\cite{Yu:2021cco}
\begin{eqnarray}
\epsilon_1^{}=\frac{3}{16\pi}\frac{I_{044}}{I_{002}\left(I_{022}-I_{002}I_{020}\right)}\;.
\end{eqnarray}
Then, via Eq.~(\ref{eq:odd6}), $\epsilon_1^{}$ can be related to the CP-odd basic flavor invariant ${\cal I}_{042}^{}$ in the SEFT. Furthermore, using four syzygies involving CP-odd invariants at degree 8, one can express ${\cal I}_{042}^{}$ as the linear combination of ${\cal I}_{121}^{(2)}$ and ${\cal I}_{240}^{}$~\cite{Yu:2022}. Finally one arrives at
\begin{eqnarray}
	\label{eq:connection}
	\epsilon_1^{}={\cal R}_1^{}\left[I_{\rm even}^{}\right]\,{\cal I}_{121}^{(2)}+{\cal R}_2^{} \left[I_{\rm even}^{}\right]\,{\cal I}_{240}^{}\;,
\end{eqnarray} 
where ${\cal R}_1^{}\left[I_{\rm even}^{}\right]$ and ${\cal R}_2^{}\left[I_{\rm even}^{}\right]$ are rational functions of only CP-even  basic invariants in the full theory that listed in Table~\ref{table:2g seesaw}. As ${\cal I}_{121}^{(2)}$ and ${\cal I}_{240}^{}$ are respectively responsible for CP violation in neutrino oscillations and neutrino-antineutrino oscillations, Eq.~(\ref{eq:connection}) establishes a direct link between low- and high-energy CP asymmetries in a basis-independent way. If ${\cal A}_{\nu\nu}^{}={\cal A}_{\nu\bar{\nu}}^{}=0$, which means ${\cal I}_{121}^{(2)}={\cal I}_{240}^{}=0$, then $\epsilon_1^{}$ also vanishes. This is obviously in accordance with the conclusion drawn from Eqs.~(\ref{eq:odd1})-(\ref{eq:odd6}) that CP conservation in the SEFT also implies the absence of CP violation in the full seesaw model.

\renewcommand\arraystretch{1.2}
\begin{table}[t!]
	\centering
	\begin{tabular}{l|c|c}
		\hline \hline
		flavor invariants &  degree & CP parity \\
		\hline \hline
		$I_{200}^{}\equiv {\rm Tr}\left(X_l^{}\right)\quad(*)$ &  2 & + \\
		\hline
		$I_{020}^{}\equiv {\rm Tr}\left(X_\nu^{}\right)\quad(*)$ &  2 & +\\
		\hline
		$I_{002}^{}\equiv {\rm Tr}\left(X_{\rm R}^{}\right)\quad(*)$ &  2 &+\\
		\hline
		$I_{400}^{}\equiv {\rm Tr}\left(X_l^2\right)\quad(*)$ & 4 &+\\
		\hline
		$I_{220}^{}\equiv {\rm Tr}\left(X_l^{}X_\nu^{}\right)\quad(*)$ & 4 &+\\
		\hline
		$I_{040}^{}\equiv {\rm Tr}\left(X_\nu^2\right)\quad(*)$ &  4 &+\\
		\hline
		$I_{022}^{}\equiv {\rm Tr}\left(\tilde{X}_\nu^{}X_{\rm R}^{}\right)\quad(*)$ &  4 & $+$\\
		\hline
		$I_{004}^{}\equiv {\rm Tr}\left(X_{\rm R}^2\right)\quad(*)$ & 4 & $+$\\
		\hline
		$I_{222}^{}\equiv {\rm Tr}\left(X_{\rm R}^{}G_{l\nu}^{}\right)\quad(*)$ & 6 & $+$\\
		\hline
		$I_{042}^{}\equiv {\rm Tr}\left(\tilde{X}_\nu^{}G_{\nu{\rm R}}^{}\right)$ & 6 & $+$\\
		\hline
		$I_{242}^{(1)}\equiv {\rm Tr}\left(G_{l\nu}^{}G_{\nu{\rm R}}^{}\right)$ & 8 & $+$\\
		\hline
		$I_{242}^{(2)}\equiv {\rm Im}\,{\rm Tr}\left(\tilde{X}_\nu^{}X_{\rm R}^{}G_{l\nu}^{}\right)$ & 8 & $-$\\
		\hline
		$I_{044}^{}\equiv {\rm Im}\,{\rm Tr}\left(\tilde{X}_\nu^{}X_{\rm R}^{}G_{\nu{\rm R}}^{}\right)$ & 8 & $-$\\
		\hline
		$I_{442}^{}\equiv {\rm Tr}\left(G_{l\nu}^{}G_{l\nu{\rm R}}^{}\right)\quad(*)$ & 10 & $+$\\
		\hline
		$I_{262}^{}\equiv {\rm Im}\,{\rm Tr}\left({\tilde X}_\nu^{}G_{l\nu}^{}G_{\nu{\rm R}}^{}\right)$ & 10 & $-$\\
		\hline
		$I_{244}^{}\equiv {\rm Im}\,{\rm Tr}\left(X_{\rm R}^{}G_{l\nu}^{}G_{\nu{\rm R}}^{}\right)$ & 10 & $-$\\
		\hline
		$I_{462}^{}\equiv {\rm Im}\,{\rm Tr}\left(\tilde{X}_\nu^{}G_{l\nu}^{}G_{l\nu{\rm R}}^{}\right)$ & 12 & $-$\\
		\hline
		$I_{444}^{}\equiv {\rm Im}\,{\rm Tr}\left(X_{\rm R}^{}G_{l\nu}^{}G_{l\nu{\rm R}}^{}\right)$ & 12 & $-$\\
		\hline
		\hline
	\end{tabular}
	\vspace{0.5cm}
	\caption{Summary of the basic flavor invariants along with their degrees and CP parities in the case of two-generation leptons in type-I seesaw model. The subscripts of the invariants denote the degrees of $Y_l^{}$, $Y_\nu^{}$ and $Y_{\rm R}^{}$, respectively. We have also defined some building blocks that transform adjointly under the flavor transformation: $X_l^{}\equiv Y_l^{}Y_l^\dagger$, $X_\nu^{}\equiv Y_\nu^{} Y_\nu^\dagger$, $\tilde{X}_\nu^{}\equiv Y_\nu^\dagger Y_\nu^{}$, $X_{\rm R}^{}\equiv Y_{\rm R}^\dagger Y_{\rm R}^{}$, $G_{l\nu}^{}\equiv Y_\nu^\dagger X_l^{} Y_\nu^{}$, $G_{\nu{\rm R}}^{}\equiv Y_{\rm R}^{\dagger} \tilde{X}_\nu^* Y_{\rm R}^{}$ and $G_{l\nu{\rm R}}^{}\equiv Y_{\rm R}^\dagger G_{l\nu}^*Y_{\rm R}^{}$. There are in total 12 CP-even basic invariants and 6 CP-odd basic invariants. The 10 primary invariants are labeled with ``$(*)$" in the first column.}
	\label{table:2g seesaw}
\end{table}
\renewcommand\arraystretch{1}

\vspace{0.3cm}

\prlsection{Three-generation Case}{.} All the results obtained in the two-generation SEFT can be generalized to the realistic three-generation scenario in a straightforward way, though the calculations are much more complicated. In this letter, we just collect the main conclusions and will present the details in a separate work~\cite{Yu:2022}. 

First, the HS in the three-generation SEFT can be computed by using the MW formula, whose expression is much lengthier than that in Eq.~(\ref{eq:HS eff 2g main}). However, as a highly nontrivial result, we find that the denominator of the HS has 21 factors, which exactly matches the number of the independent physical parameters in the SEFT. 
On the other hand, there are also 21 independent physical parameters in the three-generation seesaw. Moreover, the HS in the three-generation seesaw has been calculated in Ref.~\cite{Hanany:2010vu} and its denominator also has 21 factors.
This implies there are 21 primary invariants in both the SEFT and the full theory for three generations. Second, those 21 primary invariants in the SEFT can be explicitly constructed and from them we can extract all the physical parameters. Among them, there are 6 CP-odd invariants, corresponding to 6 independent phases in the SEFT. In particular, any CP-violating observables can also be cast into the form of Eq.~(\ref{eq:general CP asy}). Third, the vanishing of six certain CP-odd flavor invariants serves as the minimal sufficient and necessary conditions for CP conservation in the leptonic sector. The absence of CP violation in the SEFT is enough to guarantee CP conservation in the full theory, and vice versa. Finally, any flavor invariants in the SEFT can be written as rational functions of those in the full theory, which as the matching conditions set a connection between low- and high-energy observables. 
\vspace{0.3cm}

\prlsection{Concluding remarks}{.} The invariant theory is an extremely useful tool for studying CP violation in nature. Any physical observables should be independent of the flavor basis and the specific parametrization of Yukawa matrices that one chooses. This feature is exactly what flavor invariants own. Therefore it is more natural to express observables in a complete form of flavor invariants.

In this letter, we demonstrate the intimate connection between the canonical seesaw model and its low-energy effective theory in the language of invariant theory. We show that the inclusion of only one dimension-five and one dimension-six operator in the effective theory is already adequate to contain all physical information about the full theory, including the source of CP violation. The HS of the flavor space in the SEFT is calculated and all the physical parameters are explicitly extracted using primary invariants, which is helpful for phenomenological studies at low energies. The matching between flavor invariants in the SEFT and those in the full seesaw model is accomplished, offering a basis-independent way to relate CP violation for cosmological matter-antimatter asymmetry to that in low-energy phenomena.

The results in this work prove the usefulness and power of the invariant theory, and call for more applications of flavor invariants to flavor puzzles as well as other important topics in particle physics in general.

\vspace{0.3cm}

\prlsection{Acknowledgements}{.} This work was supported by the National Natural Science Foundation of China under grant No.~11835013 and the Key Research Program of the Chinese Academy of Sciences under grant No. XDPB15.

\appendix

\begin{widetext}
	\newpage
	\begin{center}
		{\bf SUPPLEMENTAL MATERIALS}
	\end{center}

In these supplemental materials, we provide the indispensable details about the results presented in the main text. First we show how to calculate the Hilbert series (HS) in the seesaw effective field theory (SEFT) using the Molien-Weyl (MW) formula. Then we demonstrate how to extract all the physical parameters in terms of primary flavor invariants. Finally we give a complete matching between the basic flavor invariants in the SEFT and those in the full theory. All the calculations are performed for the two-generation case. The generalization to the three-generation case is straightforward but much more complicated, and will be discussed in a separate work~\cite{Yu:2022}.

\section{A: Calculation of the Hilbert series in the SEFT}
\noindent A systematic method to calculate the HS is to use the MW formula~\cite{Molien1897,Weyl1926}
\begin{eqnarray}
	\label{eq:MW formula}
	{\mathscr H}(q)=\int \left[{\rm d}\mu\right]_{G}^{} {\rm PE}\left(z_1^{},...,z_{r_0}^{};q\right)\;,
\end{eqnarray}
which reduces the calculation of the HS into the computation of complex integrals. Here $\left[{\rm d}\mu\right]_G^{}$ is the Haar measure of the symmetry group $G$. The integrand is the
plethystic exponential (PE) function that determined by the representations of the building blocks
\begin{eqnarray}
	{\rm PE}\left(z_1^{},...z_{r_0}^{};q\right)={\rm exp}\left[\sum_{k=1}^{\infty}\sum_{i=1}^{n}\frac{\chi_{R_i}\left(z_1^k,...,z_{r_0}^k\right)q^k}{k}\right]\;,
\end{eqnarray}
where $z_i^{}$ (for $i=1,2,...,r_0^{}$) are coordinates on the \emph{maximum torus} of the symmetry group $G$ with $r_0^{}$ the rank of $G$. $\chi_{R_i}^{}$ (for $i=1,2,...,n$)  are the character functions for the $n$ building blocks that transform as the $R_i^{}$ representation of $G$. For the case of two-generation SEFT, the symmetry group is the two-dimensional unitary group ${\rm U}(2)$ whose rank is 2 and the character functions of the building blocks $X_l^{}\equiv Y_l^{}Y_l^\dagger$, $C_5^{}$ and $C_6^{}$ turn out to be   
\begin{eqnarray}
	\chi_l^{}\left(z_1^{},z_2^{}\right)&=&\left(z_1^{}+z_2^{}\right)\left(z_1^{-1}+z_2^{-1}\right)\;,\nonumber\\
	\chi_5^{}\left(z_1^{},z_2^{}\right)&=&z_1^2+z_2^2+z_1^{}z_2^{}+z_1^{-1}+z_2^{-1}+z_1^{-1}z_2^{-1}\;,\nonumber\\
	\chi_6^{}\left(z_1^{},z_2^{}\right)&=&\left(z_1^{}+z_2^{}\right)\left(z_1^{-1}+z_2^{-1}\right)\;.
\end{eqnarray}
Then the PE function reads	
\begin{eqnarray}
	\label{eq:PE eff 2g}
	{\rm PE}\left(z_1^{},z_2^{};q\right)&=& {\rm exp}\left(\sum_{k=1}^\infty\frac{\chi_l\left(z_1^k,z_2^k\right)q^k+\chi_5\left(z_1^k,z_2^k\right)q^k+\chi_6\left(z_1^k,z_2^k\right)q^k}{k}\right)\nonumber\\
	&=&\left[\left(1-q\right)_{}^4\left(1-qz_1^{}z_2^{-1}\right)_{}^2\left(1-qz_2^{}z_1^{-1}\right)_{}^2\left(1-qz_1^2\right)\left(1-qz_2^2\right)\left(1-qz_1^{}z_2^{}\right)\right.\nonumber\\
	&&\left.\times\left(1-qz_1^{-2}\right)\left(1-qz_2^{-2}\right)\left(1-qz_1^{-1}z_2^{-1}\right)\right]_{}^{-1}\;.
\end{eqnarray}	
Taking into account the Haar measure of ${\rm U}(2)$ group, one can calculate the HS using the MW formula in Eq.~(\ref{eq:MW formula})
\begin{eqnarray}
	\label{eq:HS eff 2g}
	{\mathscr H}_{\rm SEFT}^{(2\rm g)}(q)&=&\int \left[{\rm d}\mu\right]_{\rm U (2)}^{} {\rm PE}\left(z_1^{},z_2^{};q\right)=\frac{1}{2}\frac{1}{\left(2\pi i\right)^2}\oint_{\left|z_1\right|=1}\oint_{\left|z_2\right|=1}\left(2-\frac{z_1}{z_2}-\frac{z_2}{z_1}\right) {\rm PE}\left(z_1^{},z_2^{};q\right)\nonumber\\
	&=&\frac{1+3q^4+2q^5+3q^6+q^{10}}{\left(1-q\right)^2\left(1-q^2\right)^4\left(1-q^3\right)^2\left(1-q^4\right)^2}\;,
\end{eqnarray}
where in the final step the complex integrals are accomplished via the residue theorem. The plethystic logarithm (PL) function, which carries the information about basic invariants and syzygies~\cite{Benvenuti:2006qr}, is the inverse operation of the PE function and can be calculated by
\begin{eqnarray}
	\label{eq:PL eff 2g}
	{\rm PL}\left[{\mathscr H}_{\rm SEFT}^{(2\rm g)}(q)\right]=
	\sum_{k=1}^{\infty}\frac{\mu(k)}{k}{\rm ln}\left[{\mathscr H}_{\rm SEFT}^{(2\rm g)}(q_{}^k)\right]=2q+4q^2+2q^3+5q^4+2q^5+3q^6-6q^8-{\cal O}\left(q^{10}\right)\;,
\end{eqnarray}
where $\mu(k)$ is the M\"obius function. With the help of the positive terms in Eq.~(\ref{eq:PL eff 2g}) one can conveniently construct all the basic flavor invariants, as collected in Table~I of the main text.
Given these 18 basic flavor invariants, any flavor invariant in two-generation SEFT can be expressed as the polynomial of them. A general algorithm has been developed in Appendix C of Ref.~\cite{Wang:2021wdq} to decompose an arbitrary invariant into the polynomial of the basic ones as well as finding out all the syzygies among the basic invariants at a certain degree. Here we only list the four syzygies that involve CP-odd invariants at degree 8
\begin{eqnarray}
	\label{eq:syzygy1}
	{\cal I}_{121}^{(2)}\left(2{\cal I}_{220}^{}-{\cal I}_{100}^{}{\cal I}_{120}^{}\right)+{\cal I}_{221}^{}\left({\cal I}_{100}^{}{\cal I}_{020}^{}-2{\cal I}_{120}^{}\right)+{\cal I}_{240}^{}\left({\cal I}_{001}^{}{\cal I}_{100}^{}-2{\cal I}_{101}^{}\right) +{\cal I}_{141}^{}\left({\cal I}_{100}^2-2{\cal I}_{200}^{}\right) &=& 0\;,\\
	\label{eq:syzygy2}
	{\cal I}_{121}^{(2)}\left(2{\cal I}_{022}^{}-{\cal I}_{001}^{}{\cal I}_{021}^{}\right)-{\cal I}_{122}^{}\left({\cal I}_{001}^{}{\cal I}_{020}^{}-2{\cal I}_{021}^{}\right)-{\cal I}_{042}^{}\left({\cal I}_{001}^{}{\cal I}_{100}^{}-2{\cal I}_{101}^{}\right)-{\cal I}_{141}^{}\left({\cal I}_{001}^2-2{\cal I}_{002}^{}\right) &=& 0\;,\\
	\label{eq:syzygy3}
	{\cal I}_{121}^{(2)}\left(2{\cal I}_{121}^{(1)}-{\cal I}_{001}^{}{\cal I}_{120}^{}\right)+{\cal I}_{221}^{}\left({\cal I}_{001}^{}{\cal I}_{020}^{}-2{\cal I}_{021}^{}\right)+{\cal I}_{240}^{}\left({\cal I}_{001}^2-2{\cal I}_{002}^{}\right)+{\cal I}_{141}^{}\left({\cal I}_{001}^{}{\cal I}_{100}^{}-2{\cal I}_{101}^{}\right) &=& 0\;,\\
	\label{eq:syzygy4}
	{\cal I}_{121}^{(2)}\left(2{\cal I}_{121}^{(1)} -{\cal I}_{021}^{}{\cal I}_{100}^{}\right)-{\cal I}_{122}^{}\left({\cal I}_{020}^{}{\cal I}_{100}^{}-2{\cal I}_{120}^{}\right)-{\cal I}_{042}^{}\left({\cal I}_{100}^2-2{\cal I}_{200}^{}\right)-{\cal I}_{141}^{}\left({\cal I}_{001}^{}{\cal I}_{100}^{}-2{\cal I}_{101}^{}\right) &=& 0\;,
\end{eqnarray}
from which one can express any four of the six CP-odd basic invariants in Table~I of the main text as the linear combinations of the other two, with the coefficients being rational functions of only CP-even basic invariants.

The HS in the full seesaw model can be computed using the same method
\begin{eqnarray}
	\label{eq:HS seesaw 2g}
	{\mathscr H}_{\rm SS}^{(2\rm g)}(q)
	&=&\int \left[{\rm d}\mu\right]_{{\rm U}(2)\otimes{\rm U}(2)}^{} {\rm PE}\left(z_1^{},z_2^{},z_3^{},z_4^{};q\right)\nonumber\\
	&=&\frac{1}{4}\frac{1}{\left(2\pi {\rm i}\right)^4}\oint_{\left|z_1\right|=1}\oint_{\left|z_2\right|=1}\oint_{\left|z_3\right|=1}\oint_{\left|z_4\right|=1}\left(2-\frac{z_1}{z_2}-\frac{z_2}{z_1}\right)\left(2-\frac{z_3}{z_4}-\frac{z_4}{z_3}\right) {\rm PE}\left(z_1^{},z_2^{},z_3^{},z_4^{};q\right)\nonumber\\
	&=&\frac{1+q^6+3q^8+2q^{10}+3q^{12}+q^{14}+q^{20}}{\left(1-q^2\right)^3\left(1-q^4\right)^5\left(1-q^6\right)\left(1-q^{10}\right)}\;,
\end{eqnarray}
while the PL function turns out to be
\begin{eqnarray}
	\label{eq:PL seesaw 2g}
	{\rm PL}\left[{\mathscr H}_{\rm SS}^{(2\rm g)}(q)\right]=3q^2+5q^4+2q^6+3q^8+3q^{10}+2q^{12}-{\cal O}\left(q^{14}\right)\;.
\end{eqnarray}
From the positive terms in Eq.~(\ref{eq:PL seesaw 2g}) one can read off that there are also 18 basic invariants in the invariant ring of the full seesaw model: three of degree 2, five of degree 4, two of degree 6, three of degree 8, three of degree 10 and two of degree 12. With the help of Eq.~(\ref{eq:PL seesaw 2g}) we have explicitly constructed all the basic invariants in the full seesaw model, as shown in Table II of the main text.

\section{B: Physical parameters in terms of flavor invariants}
\noindent In the flavor basis where $C_5^{}$ is diagonal with real and positive eigenvalues, i.e., $C_5^{}={\rm Diag}\{c_1^{},c_2^{}\}$, one can generally write the $2\times 2$ Hermitian matrices $X_l^{}\equiv Y_l^{} Y_l^\dagger$ and $C_6^{}$ as follows
\begin{eqnarray}
	\label{eq:parametrization of C6 2g}
	X_l^{}=\left(
	\begin{matrix}
		a_{11}^{}&a_{12}^{}e_{}^{i\alpha}\\
		a_{12}^{}e_{}^{-i\alpha}&a_{22}^{}
	\end{matrix}
	\right)\;,
	C_6^{}=\left(
	\begin{matrix}
		b_{11}^{}&b_{12}^{}e_{}^{i\beta}\\
		b_{12}^{}e_{}^{-i\beta}&b_{22}^{}
	\end{matrix}
	\right)\;,
\end{eqnarray}
where $a_{ij}^{}$ and $b_{ij}^{}$ are real numbers while $\alpha$ and $\beta$ are two phases. In this basis, 10 independent physical parameters are collected as $\{c_1^{},c_2^{},a_{11}^{},a_{12}^{},a_{22}^{},b_{11}^{},b_{12}^{},b_{22}^{},\alpha,\beta\}$.

Now we extract these ten parameters from the primary invariants. First, the eigenvalues of $C_5^{}$ can be obtained from ${\cal I}_{020}^{}\equiv {\rm Tr}\left(X_5^{}\right)$ and ${\cal I}_{040}^{}\equiv {\rm Tr}\left(X_5^2\right)$ with $X^{}_5 \equiv C^{}_5 C^\dagger_5$  as below
\begin{eqnarray}
	c_{1,2}^{}=\frac{1}{\sqrt{2}}\sqrt{{\cal I}_{020}^{}\mp\sqrt{2{\cal I}_{040}^{}-{\cal I}_{020}^2}}\; ,
\end{eqnarray}
where $c^{}_{1}$ and $c^{}_2$ corresponds to the upper and lower sign on the right-hand side, respectively.
Then, from ${\cal I}_{100}^{}\equiv {\rm Tr}\left(X_l^{}\right)=a_{11}^{}+a_{22}^{}$ and ${\cal I}_{120}^{}\equiv {\rm Tr}\left(X_l^{}X_5^{}\right)=c_1^2a_{11}^{}+c_2^2a_{22}^{}$ one can find
\begin{eqnarray}
	a_{11,22}^{}=\frac{1}{2}\left({\cal I}_{100}^{}\pm\frac{{\cal I}_{100}{\cal I}_{020}-2{\cal I}_{120}}{\sqrt{2{\cal I}_{040}-{\cal I}_{020}^2}}\right)\;.
\end{eqnarray}
With the help of ${\cal I}_{200}^{}\equiv {\rm Tr}\left(X_l^2\right)=a_{11}^2+2a_{12}^2+a_{22}^2$, we can immediately get
\begin{eqnarray}
	a_{12}^{}=\frac{1}{\sqrt{2}}\sqrt{\frac{{\cal I}_{100}\left({\cal I}_{100}{\cal I}_{040}-2{\cal I}_{020}{\cal I}_{120}\right)+{\cal I}_{200}\left({\cal I}_{020}^2-2{\cal I}_{040}\right)+2{\cal I}_{120}^2}{{\cal I}_{020}^2-2{\cal I}_{040}}}\;.
\end{eqnarray}
Finally, using the identity ${\cal I}_{220}^{}\equiv {\rm Tr}\left(X_l^{}G_{l5}^{}\right)=c_1^2 a_{11}^2+c_2^2 a_{22}^2+2a_{12}^2c_1^{}c_2^{}\cos2\alpha$ with $G_{l5}^{}\equiv C_5^{}X_l^*C_5^\dagger$, one can solve $\cos2\alpha$ in terms of primary invariants, i.e.,
\begin{eqnarray}
	\label{eq:extractalpha}
	\cos2\alpha=\frac{\left({\cal I}_{100}^2{\cal I}_{020}-4{\cal I}_{100}{\cal I}_{120}+2{\cal I}_{220}\right)\left({\cal I}_{020}^2-{\cal I}_{040}\right)+2\left({\cal I}_{020}{\cal I}_{120}^2-{\cal I}_{040}{\cal I}_{220}\right)}{\sqrt{2}\sqrt{{\cal I}_{020}^2-{\cal I}_{040}}\left[{\cal I}_{200}^{}\left({\cal I}_{020}^2-{\cal I}_{040}\right)+{\cal I}_{040}\left({\cal I}_{100}^2-{\cal I}_{200}\right)-2{\cal I}_{120}\left({\cal I}_{100}{\cal I}_{020}-{\cal I}_{120}\right)\right]}\;.
\end{eqnarray}
In a similar way, four real parameters in $C_6^{}$ can be determined by 
\begin{eqnarray}
	\label{eq:extract C6 2g 1}
	b_{11,22}^{}&=&\frac{1}{2}\left({\cal I}_{001}^{}\pm\frac{{\cal I}_{001}{\cal I}_{020}-2{\cal I}_{021}}{\sqrt{2{\cal I}_{040}-{\cal I}_{020}^2}}\right)\;,\\
	\label{eq:extract C6 2g 2}
	b_{12}^{}&=&\frac{1}{\sqrt{2}}\sqrt{\frac{{\cal I}_{001}\left({\cal I}_{001}{\cal I}_{040}-2{\cal I}_{020}{\cal I}_{021}\right)+{\cal I}_{002}\left({\cal I}_{020}^2-2{\cal I}_{040}\right)+2{\cal I}_{021}^2}{{\cal I}_{020}^2-2{\cal I}_{040}}}\;,\\
	\label{eq:extract C6 2g 3}
	\cos2\beta&=&\frac{\left({\cal I}_{001}^2{\cal I}_{020}-4{\cal I}_{001}{\cal I}_{021}+2{\cal I}_{022}\right)\left({\cal I}_{020}^2-{\cal I}_{040}\right)+2\left({\cal I}_{020}{\cal I}_{021}^2-{\cal I}_{040}{\cal I}_{022}\right)}{\sqrt{2}\sqrt{{\cal I}_{020}^2-{\cal I}_{040}}\left[{\cal I}_{002}^{}\left({\cal I}_{020}^2-{\cal I}_{040}\right)+{\cal I}_{040}\left({\cal I}_{001}^2-{\cal I}_{002}\right)-2{\cal I}_{021}\left({\cal I}_{001}{\cal I}_{020}-{\cal I}_{021}\right)\right]}\;.
\end{eqnarray}
After the spontaneous breakdown of the SM gauge symmetry, the neutrino masses are just given by the eigenvalues of $C_5^{}$ via
\begin{eqnarray}
	\label{eq:extract neutrino mass 2g}
	m_{1,2}^{}=\frac{v^2}{2\Lambda}c_{1,2}^{}=\frac{v^2}{2\sqrt{2}\Lambda}\sqrt{{\cal I}_{020}^{}\mp\sqrt{2{\cal I}_{040}^{}-{\cal I}_{020}^2}}\;. \quad
\end{eqnarray}

Here we would like to give some comments on the equivalence between the set of physical observables and the set of primary invariants. Strictly speaking, there may be discrete leftover degeneracies when extracting physical parameters by using primary invariants. For example, in Eqs.~(\ref{eq:extractalpha}) and (\ref{eq:extract C6 2g 3}) the primary invariants are blind to the signs of $\alpha$ and $\beta$. The degeneracies can be eliminated by including more basic CP-odd invariants other than the primary ones. In our case, one can introduce ${\cal I}_{240}\propto \sin2\alpha$ and ${\cal I}_{042}\propto \sin2\beta$ (note that ${\cal I}_{240}$ and ${\cal I}_{042}$ are \emph{not} algebraic independent of the primary invariants), whose signs can be used to eliminate the $Z_2$ degeneracy $\alpha\to -\alpha$ and $\beta\to -\beta$. Therefore, a more strict statement should be: ``The set of physical
parameters is equivalent to the set of primary invariants in the ring \emph{up to some discrete degeneracies}''~\cite{Yu:2022}.        

On the other hand, the charged-lepton masses can be figured out from $X_l^{}$ via $2\,{\rm Diag}\left\{m_e^2,m_\mu^2\right\}/v^2=V_2^{}X_l^{}V_2^\dagger$, where the $2\times 2$ flavor mixing matrix is parametrized as
\begin{eqnarray}
	\label{eq:parametrization of V 2g}
	V_2^{}=\left(
	\begin{matrix}
		\cos\theta&\sin\theta\\
		-\sin\theta&\cos\theta
	\end{matrix}
	\right)\cdot
	\left(
	\begin{matrix}
		e_{}^{i\phi}&0\\
		0&1
	\end{matrix}
	\right)\;,
\end{eqnarray} 
with $\theta$ being the flavor mixing angle and $\phi$ being the Majorana-type CP phase. It is straightforward to relate the charged-lepton masses, the flavor mixing angle and the CP phase to the elements in $X_l^{}$ by $\phi = -\alpha$ and 
\begin{eqnarray}
	\tan2\theta &=& \frac{2a_{12}}{a_{11}-a_{22}}\;, \\
	m_{e,\mu^{}} &=& \frac{v}{2}\sqrt{a_{11}^{}+a_{22}^{}\pm\frac{2a_{12}^{}}{\sin2\theta}}\;.
\end{eqnarray}
More explicitly, we can express these parameters in terms of the primary flavor invariants, i.e.,
\begin{eqnarray}
	\label{eq:extract chargd-lepton mass 2g}
	m_{e,\mu}&=&\frac{v}{2}\sqrt{{\cal I}_{100}^{}\mp\sqrt{2{\cal I}_{200}^{}-{\cal I}_{100}^2}}\;,\\
	\label{eq:extract theta}
	\cos2\theta&=&\frac{2{\cal I}_{120}-{\cal I}_{100}{\cal I}_{020}}{\sqrt{2{\cal I}_{040}-{\cal I}_{020}^2}\sqrt{2{\cal I}_{200}-{\cal I}_{100}^2}}\;,\\
	\label{eq:extract phi}
	\cos2\phi&=&\frac{\left({\cal I}_{100}^2{\cal I}_{020}-4{\cal I}_{100}{\cal I}_{120}+2{\cal I}_{220}\right)\left({\cal I}_{020}^2-{\cal I}_{040}\right)+2\left({\cal I}_{020}{\cal I}_{120}^2-{\cal I}_{040}{\cal I}_{220}\right)}{\sqrt{2}\sqrt{{\cal I}_{020}^2-{\cal I}_{040}}\left[{\cal I}_{200}^{}\left({\cal I}_{020}^2-{\cal I}_{040}\right)+{\cal I}_{040}\left({\cal I}_{100}^2-{\cal I}_{200}\right)-2{\cal I}_{120}\left({\cal I}_{100}{\cal I}_{020}-{\cal I}_{120}\right)\right]}\;.
\end{eqnarray}
To sum up, 10 physical parameters after the gauge symmetry breaking $\left\{ m_1^{}, m_2^{}, m_e^{}, m_\mu^{}, \theta, \phi, b_{11}^{}, b_{12}^{}, b_{22}^{}, \beta \right\}$
can be extracted from the 10 primary flavor invariants 
$$\left\{{\cal I}_{020}^{},{\cal I}_{040}^{}, {\cal I}_{100}^{}, {\cal I}_{001}^{}, {\cal I}_{200}, {\cal I}_{002}^{}, {\cal I}_{120}^{}, {\cal I}_{021}^{}, {\cal I}_{220}^{}, {\cal I}_{022}^{} \right\}$$
by Eqs.~(\ref{eq:extract C6 2g 1})-(\ref{eq:extract C6 2g 3}), Eq.~(\ref{eq:extract neutrino mass 2g}) and Eqs.~(\ref{eq:extract chargd-lepton mass 2g})-(\ref{eq:extract phi}). On this point, it is worthwhile to stress that the physical parameters are certainly functions, but not polynomial functions, of primary flavor invariants. 

Therefore, any physical observables can be expressed in explicit basis-independent forms with flavor invariants. In particular, any CP-violating observable ${\cal A}_{\rm CP}^{}$ can be written as
\begin{eqnarray}
	\label{eq:general CP asy}
	{\cal A}_{\rm CP}^{}=\sum_j^{} {\cal F}_j^{}\left[{\cal I}_k^{\rm even}\right] {\cal I}_j^{\rm odd}\;,
\end{eqnarray}
where ${\cal I}_j^{\rm odd}$ refer to CP-odd basic flavor invariants, and ${\cal F}_j^{}\left[{\cal I}_k^{\rm even}\right]$ are some functions of only CP-even basic flavor invariants. For illustration, we calculate the CP asymmetries in two-generation neutrino oscillations and neutrino-antineutrino oscillations defined as  
\begin{eqnarray}
	\label{eq:CP asymmetry in neutrino oscillation def}
	{\cal A}_{\nu\nu}^{\alpha\beta}\equiv \frac{{\rm P}\left(\nu_\alpha\to\nu_\beta\right)-{\rm P}\left(\bar{\nu}_\alpha\to\bar{\nu}_\beta\right)}{{\rm P}\left(\nu_\alpha\to\nu_\beta\right)+{\rm P}\left(\bar{\nu}_\alpha\to\bar{\nu}_\beta\right)}\;,\qquad
	{\cal A}_{\nu\bar{\nu}}^{\alpha\beta}\equiv \frac{{\rm P}\left(\nu_\alpha\to \bar{\nu}_\beta\right)-{\rm P}\left(\bar{\nu}_\alpha \to \nu_\beta\right)}{{\rm P}\left(\nu_\alpha\to \bar{\nu}_{\beta}\right)+{\rm P}\left(\bar{\nu}_\alpha\to \nu_\beta\right)}\;,
\end{eqnarray}
where $\alpha,\beta=e,\mu$ and P denotes the oscillation probability of corresponding channel. The CP asymmetries can first be calculated in the mass eigenstates. Then, using above results, we are able to recast them into the form of Eq.~(\ref{eq:general CP asy})
\begin{eqnarray}
	A_{\nu\nu}^{e\mu}&=&-\frac{v^2}{\Lambda^2}\cot\left(\frac{\Delta_{21}}{2}\right){\cal F}_{\nu\nu}^{e\mu}\left[{\cal I}_{100}^{},{\cal I}_{200}^{},{\cal I}_{020}^{},{\cal I}_{120}^{},{\cal I}_{040}^{}\right]\,{\cal I}_{121}^{(2)}\;,\nonumber\\
	{\cal A}_{\nu\bar{\nu}}^{e\mu}&=&{\cal F}_{\nu\bar{\nu}}^{e\mu}\left[{\cal I}_{100}^{},{\cal I}_{200}^{},{\cal I}_{020}^{},{\cal I}_{120}^{},{\cal I}_{220}^{},{\cal I}_{040}^{}\right]\,{\cal I}_{240}^{}\;,
\end{eqnarray}
where 
\begin{eqnarray}
	{\cal F}_{\nu\nu}^{e\mu}\left[{\cal I}_{100}^{},{\cal I}_{200}^{},{\cal I}_{020}^{},{\cal I}_{120}^{},{\cal I}_{040}^{}\right]=\frac{\left(2{\cal I}_{040}-{\cal I}_{020}^2\right)^{1/2}\left(2{\cal I}_{200}-{\cal I}_{100}^2\right)^{1/2}}{{\cal I}_{040}\left(2{\cal I}_{200}-{\cal I}_{100}^2\right)-2{\cal I}_{120}^{}\left({\cal I}_{120}-{\cal I}_{020}{\cal I}_{100}\right)-{\cal I}_{020}^2{\cal I}_{200}}\;,
\end{eqnarray}
and
\begin{eqnarray}
	&&{\cal F}_{\nu\bar{\nu}}^{e\mu}\left[{\cal I}_{100}^{},{\cal I}_{200}^{},{\cal I}_{020}^{},{\cal I}_{120}^{},{\cal I}_{220}^{},{\cal I}_{040}^{}\right]\nonumber\\
	&=&4\left(2{\cal I}_{040}^{}-{\cal I}_{020}^2\right)_{}^{1/2}\sin\Delta_{21}^{}
	\left\{{\cal I}_{020}^{}\left[2{\cal I}_{120}^{}\left({\cal I}_{100}^{}{\cal I}_{020}^{}-{\cal I}_{120}^{}\right)-{\cal I}_{040}^{}\left({\cal I}_{100}^2-2{\cal I}_{200}^{}\right)-{\cal I}_{020}^2{\cal I}_{200}^{}\right]\right.\nonumber\\
	&&\left.+\cos\Delta_{21}^{}\left[{\cal I}_{020}^{}\left({\cal I}_{020}^2{\cal I}_{100}^2+2{\cal I}_{120}^2-{\cal I}_{040}^{}{\cal I}_{100}^2\right)+4{\cal I}_{040}^{}\left({\cal I}_{100}^{}{\cal I}_{120}^{}-{\cal I}_{220}^{}\right)+2{\cal I}_{020}^2\right.\right.\nonumber\\
	&&\left.\left.\times\left({\cal I}_{220}^{}-2{\cal I}_{100}^{}{\cal I}_{120}^{}\right)\right]\right\}_{}^{-1}\;,
\end{eqnarray}
with 
\begin{eqnarray}
	\Delta_{21}^{}= \frac{L}{2E}\left(m_2^2-m_1^2\right)=\frac{Lv^4}{8E\Lambda^2}\left(2{\cal I}_{040}^{}-{\cal I}_{020}^2\right)_{}^{1/2}\;,
\end{eqnarray}
where $L$ and $E$ denote the propagation distance and neutrino beam energy, respectively. It is obvious that ${\cal A}_{\nu\nu}^{e\mu}$ is suppressed by $v_{}^2/\Lambda_{}^2$, since there is no CP violation in two-generation neutrino oscillations without including $C_6^{}$. However, ${\cal A}_{\nu\bar{\nu}}^{e\mu}$ is not suppressed, as the Majorana-type CP phase in the $2\times 2$ leptonic flavor mixing matrix already enters into the CP asymmetries in neutrino-antineutrino oscillations.

\section{C: Flavor invariants matching}
\noindent Here we list the complete set of matching conditions between the basic invariants in the SEFT and those in the full seesaw model. First, the 12 CP-even basic flavor invariants in the SEFT can be written as the rational functions of the 12 CP-even basic flavor invariants in the full theory
{\allowdisplaybreaks
	\begin{eqnarray}
		\label{eq:even1app}
		{\cal I}_{100}^{}&=&I_{200}^{}\;,\\
		{\cal I}_{001}^{}&=&\frac{2}{\left(I_{002}^2-I_{004}\right)}\left(I_{002}^{}I_{020}^{}-I_{022}^{}\right)\;,\\
		{\cal I}_{200}^{}&=&I_{400}^{}\;,\\
		{\cal I}_{101}^{}&=&\frac{2}{\left(I_{002}^2-I_{004}\right)}\left(I_{002}^{}I_{220}^{}-I_{222}^{}\right)\;,\\
		{\cal I}_{020}^{}&=&\frac{2}{\left(I_{002}^2-I_{004}\right)}\left(I_{042}^{}-2I_{022}^{}I_{020}^{}+I_{020}^2I_{002}^{}\right)\;,\\
		{\cal I}_{002}^{}&=&\frac{2}{\left(I_{002}^2-I_{004}\right)^2}\left[2I_{022}^{}\left(I_{022}^{}-2I_{002}^{}I_{020}^{}\right)+I_{004}^{}\left(I_{020}^2-I_{040}^{}\right)+I_{002}^2\left(I_{020}^2+I_{040}^{}\right)\right]\;,\\
		{\cal I}_{120}^{}&=&\frac{2}{\left(I_{002}^2-I_{004}\right)}\left[I_{220}^{}\left(I_{020}^{}I_{002}^{}-I_{022}^{}\right)+I_{242}^{(1)}-I_{020}^{}I_{222}^{}\right]\;,\\
		{\cal I}_{021}^{}&=&\frac{1}{\left(I_{002}^2-I_{004}\right)^2}\left[I_{004}^{}I_{020}^{}\left(I_{020}^2-I_{040}^{}\right)+I_{002}^2I_{020}^{}\left(3I_{020}^2+I_{040}^{}\right)-4I_{022}^{}\left(I_{042}^{}-2I_{020}^{}I_{022}^{}\right)\right.\nonumber\\
		&&\left.+4I_{002}^{}I_{020}^{}\left(I_{042}^{}-3I_{020}^{}I_{022}^{}\right)\right]\;,\\
		{\cal I}_{220}^{}&=&\frac{2}{\left(I_{002}^2-I_{004}\right)}\left[I_{220}^{}\left(I_{220}^{}I_{002}^{}-2I_{222}^{}\right)+I_{442}^{}\right]\;,\\
		{\cal I}_{121}^{(1)}&=&\frac{1}{\left(I_{002}^2-I_{004}\right)^2}\left[I_{004}^{}I_{220}^{}\left(I_{020}^2-I_{040}^{}\right)+I_{002}^2I_{220}^{}\left(3I_{020}^2+I_{040}^{}\right)\right.\nonumber\\
		&&\left.+4I_{022}^{}\left(I_{022}^{}I_{220}^{}+I_{020}^{}I_{222}^{}-I_{242}^{(1)}\right)-4I_{002}^{}I_{020}^{}\left(2I_{022}^{}I_{220}^{}+I_{020}^{}I_{222}^{}-I_{242}^{(1)}\right)\right]\;,\\
		{\cal I}_{040}^{}&=&\frac{1}{\left(I_{002}^2-I_{004}\right)^2}\left[I_{004}^{}\left(I_{020}^2-I_{040}^{}\right)_{}^2+I_{002}^2\left(3I_{020}^2-I_{040}^{}\right)\left(I_{020}^2+I_{040}^{}\right)\right.\nonumber\\
		&&\left.-4\left(2I_{020}^{}I_{022}^{}-I_{042}^{}\right)\left(2I_{002}^{}I_{020}^2-2I_{020}^{}I_{022}^{}+I_{042}^{}\right)
		\right]\;,\\
		{\cal I}_{022}^{}&=&\frac{1}{\left(I_{002}^2-I_{004}\right)^3}\left[I_{002}^3\left(5I_{020}^4+2I_{020}^2I_{040}+I_{040}^2\right)+8I_{020}^{}\left(I_{002}^2I_{020}^{}I_{042}^{}-2I_{022}^3\right)\right.\nonumber\\
		&&\left.+8I_{022}^2\left(5I_{002}^{}I_{020}^2+I_{042}\right)-4I_{022}^{}I_{002}^{}I_{020}^{}\left(7I_{002}^{}I_{020}^2+I_{002}^{}I_{040}^{}+4I_{042}^{}\right)\right.\nonumber\\
		\label{eq:even12app}
		&&\left.+I_{004}^{}\left(I_{020}^2-I_{040}^{}\right)\left(3I_{002}^{}I_{020}^2+I_{002}^{}I_{040}^{}-4I_{020}^{}I_{022}^{}\right)\right]\;,
	\end{eqnarray}
	which are independent of any CP-odd invariants. Then, the 6 CP-odd basic flavor invariants in the SEFT can be written as the \emph{linear} combinations of the 6 CP-odd basic flavor invariants in the full theory, with the coefficients being rational functions of CP-even basic flavor invariants in the full theory
	\begin{eqnarray}
		{\cal I}_{121}^{(2)}&=&\frac{2}{\left(I_{002}^2-I_{004}\right)^2}\left[I_{242}^{(2)}I_{022}^{}-I_{044}^{}I_{220}^{}+I_{262}^{}I_{002}^{}-I_{244}^{}I_{020}^{}\right]\;,\label{eq:odd1app}\\
		{\cal I}_{221}^{}&=&\frac{2}{\left(I_{002}^2-I_{004}\right)^2}\left[I_{242}^{(2)}I_{222}^{}+I_{244}^{}I_{220}^{}+I_{462}^{}I_{002}^{}-I_{444}^{}I_{020}^{}\right]\;,\label{eq:odd2app}\\
		{\cal I}_{122}^{}&=&\frac{2}{\left(I_{002}^2-I_{004}\right)^3}\left\{I_{242}^{(2)}\left[3I_{022}^2+2I_{040}^{}\left(I_{002}^2-I_{004}^{}\right)-4I_{020}^{}I_{002}^{}I_{022}^{}\right] +I_{244}^{}\left(3I_{020}^{}I_{022}^{}-2I_{042}^{}\right) \right.\nonumber\\
		&&\left.+I_{044}^{}\left(4I_{020}^{}I_{222}^{}-I_{220}^{}I_{022}^{}-2I_{242}^{(1)}\right)+I_{262}^{}\left[3I_{002}^{}I_{022}^{}-I_{020}^{}\left(I_{002}^2+3I_{004}^{}\right)\right]\right\} \; , \label{eq:odd3app} \\
		{\cal I}_{240}^{}&=&\frac{1}{\left(I_{002}^2-I_{004}\right)^2}\left[3I_{242}^{(2)}\left(I_{022}^{}I_{220}^{}-I_{020}^{}I_{222}^{}\right)-I_{044}^{}I_{220}^2+I_{262}^{}\left(3I_{002}^{}I_{220}^{}-2I_{222}^{}\right) -2 I_{244}^{}I_{020}^{}I_{220}^{}\right.\nonumber\\
		&&\left.+I_{462}^{}\left(2I_{022}^{}-3I_{002}^{}I_{020}\right)+I_{444}^{}I_{020}^2\right]\;,\label{eq:odd4app}\\
		{\cal I}_{141}^{}&=&\frac{2}{\left(I_{002}^2-I_{004}\right)^3}\left\{I_{242}^{(2)}I_{020}^{}I_{022}^2+I_{044}^{}I_{020}^{}\left(I_{022}^{}I_{220}^{}-2I_{242}^{(1)}\right) +I_{244}^{}I_{020}^{}\left(I_{020}^{}I_{022}-2I_{042}^{}\right) \right.\nonumber\\
		&&\left.+I_{262}^{}\left[I_{002}^{}I_{020}^{}I_{022}^{}+I_{040}^{}\left(I_{004}^{}-I_{002}^2\right)\right]\right\}\;,\label{eq:odd5app}\\
		{\cal I}_{042}^{}&=&\frac{2}{\left(I_{002}^2-I_{004}\right)^3}\,I_{044}^{}\left(I_{020}^2-I_{040}^{}\right)_{}^2\;.\label{eq:odd6app}
	\end{eqnarray}
}
In addition,  the determinant of the coefficient matrix in Eqs.~(\ref{eq:odd1app})-(\ref{eq:odd6app}) reads
\begin{eqnarray}
	\label{eq:det}
	{\rm Det}&=&\frac{128}{\left(I_{002}^2-I_{004}^{}\right)^{14}}\,I_{020}^{}\left(I_{002}^{}I_{020}^{}-I_{022}^{}\right)\left(I_{020}^2-I_{040}^{}\right)_{}^2\nonumber\\
	&&\times\left\{I_{020}^2I_{022}^{}\left(3I_{020}^{}I_{022}^{}-4I_{002}^{}I_{040}^{}-3I_{042}^{}\right)-I_{022}^{}I_{040}^{}I_{042}^{}\right.\nonumber\\
	&&\left.+ I_{020}^{}I_{040}^{}\left[3I_{022}^2+2I_{002}^{}I_{042}^{}+I_{040}^{}\left(I_{002}^2-I_{004}^{}\right)\right]\right\}\;,
\end{eqnarray}
which is nonzero in general. This means that Eqs.~(\ref{eq:odd1app})-(\ref{eq:odd6app}) are linearly independent.

\end{widetext}
\end{document}